\documentclass[12pt,preprint]{aastex}
\usepackage{epsfig}
\usepackage{natbib}                % To format bibliographies.
\usepackage{lscape}

\shorttitle{Angular Diameters of Hyades Giants}
\shortauthors{Boyajian \& McAlister et al.}
\begin{document}

%%\received{}
%%\accepted{}

\title{Angular Diameters of the Hyades Giants Measured with the CHARA Array} 

\author{Tabetha S. Boyajian, Harold A. McAlister, Justin R. Cantrell, Douglas R. Gies}

\affil{Center for High Angular Resolution Astronomy and \\
Department of Physics and Astronomy,\\
Georgia State University, P. O. Box 4106, Atlanta, GA 30302-4106} 
\email{tabetha@chara.gsu.edu, hal@chara.gsu.edu, cantrell@chara.gsu.edu, gies@chara.gsu.edu}

\author{Theo A. ten Brummelaar, Chris Farrington, P. J. Goldfinger,  \\ Laszlo Sturmann, Judit Sturmann, Nils H. Turner}
\affil{The CHARA Array, Mount Wilson Observatory, Mount Wilson, CA 91023}
\email{theo@chara-array.org, farrington@chara-array.org, pj@chara-array.org, sturmann@chara-array.org, judit@chara-array.org, nils@chara-array.org}
\author{Stephen Ridgway}
\affil{National Optical Astronomy Observatory, P.O. Box 26732, Tucson, AZ 85726-6732}
\email{sridgway@noao.edu}
%%\slugcomment{}
%%\paperid{}

%%%%%%%%%%%%%%%%%%%%%%%%%%%%%%%%%%%%%%%%%%%%%%%%%%%%%%%%%%%%%%

\begin{abstract}

We present angular diameters of the Hyades giants, $\gamma$, $\delta^{1}$, $\epsilon$, and $\theta^{1}$~Tau from interferometric measurements with the CHARA Array.  Our errors in the limb-darkened angular diameters for these stars are all less than $2\%$, and in combination with additional observable quantities, we determine the effective temperatures, linear radii and absolute luminosities for each of these stars.  Additionally, stellar masses are inferred from model isochrones to determine the surface gravities.  These data show that a new calibration of effective temperatures with errors well under 100K is now possible from interferometric angular diameters of stars.

\end{abstract}


%%%%%%%%%%%%%%%%%%%%%%%%%%%%%%%%%%%%%%%%%%%%%%%%%%%%%%%%%%%%%%%
%    Section 1

\section{Introduction}                              

Because of its close proximity to the Sun, the Hyades cluster has served as a benchmark in studies ranging from stellar evolutionary modeling to calibrating the cosmic distance scale.  In the context of evolutionary theory, {\it Hipparcos} distances and resolved binaries in the cluster have enabled us to test extensively those models (for example, see \citealt{per98, las99}) using fundamental stellar properties such as effective temperature. The only direct way to determine the effective temperature of a star is to measure the star's angular diameter and integrated flux.  While the dwarf stars in the Hyades are too small to resolve their angular diameters with current tools and methods, the four Hyades giants have been observed over the past few decades, beginning with lunar occultation measurements (see Table~5 for references and timeline of publications of this topic).  Presently, long-baseline optical interferometry (LBOI) has trumped lunar occultation (LO) techniques in accurately measuring the angular diameters of such stars.  In fact, for the Hyades giants in particular, the accuracy in the angular diameter measurements has improved by almost an order of magnitude over the past few decades.

In this work, we present the first uniform analysis of all four of the Hyades giants, $\gamma$~Tau (HR~1346, HD~27371, HIP~20205), $\delta^{1}$~Tau (HR~1373, HD~27697, HIP~20455), $\epsilon$~Tau (HR~1409, HD~28305, HIP~20889), and $\theta^{1}$~Tau (HR~1411, HD~28307, HIP~20885).  We observed these stars with the CHARA Array to obtain their angular diameters to better than 2\% accuracy.  In combination with the bolometric flux of each star, we derive their effective temperatures to 1\% accuracy (\S3).  In this paper, we describe our observational results and then compare them to model isochrones for the Hyades cluster, which demonstrate remarkable agreement within the temperature-luminosity plane for the cluster turn-off age and metallicity (\S4).

%%%%%%%%%%%%%%%%%%%%%%%%%%%%%%%%%%%%%%%%%%%%%%%%%%%%%%%%%%%%%%%
    % Section 2

\section{Observations and Data Reduction}  

We observed these stars with the CHARA Array, located on the grounds of Mount Wilson Observatory, using the CHARA Classic beam combiner in $K^{\prime}$-band \citep{ttb05} with the W2-E2 baseline (maximum baseline of 156.3~m) on 2007 November 02 from the Georgia State University AROC\footnote{Arrington Remote Operations Center} facility in Atlanta, GA.  The chosen calibrator star, $\delta^{2}$~Tau (HR~1380, HD~27819), an A7V with $v \sin i = 42$ km s$^{-1}$ \citep{roy07}, is separated by less than two degrees on the sky for all of the targets.  It was observed in bracketed sequences with each of the target stars yielding a total of 9 bracketed observations for $\gamma$, $\delta^{1}$ and $\epsilon$~Tau, and 8 bracketed observations for $\theta^{1}$~Tau.  The angular diameter $\theta_{\rm SED}$ of the calibrator star was calculated by fitting observed photometry (see \citealt{boy08} for details) to a model spectral energy distribution (SED)\footnote{The model fluxes were interpolated from the grid of models from R. L. Kurucz available at http://kurucz.cfa.harvard.edu/}.  The close proximity of the Hyades Cluster members to us ($\sim 47$ parsecs; \citealt{van07}) introduces no effects on the SED fit due to interstellar reddening ($E(B-V)\leq 0.001$~mmag, \citealt{tay06}, and references therein). The SED model fit for the calibrator star yields $\theta_{\rm SED}=0.457\pm0.020$~mas, for an effective temperature of $T_{\rm eff}$=8100~K and $\log g=4.1$.  This corresponds to an absolute calibrated visibility for the calibrator star of $\sim0.97$ at these baselines.  Data reduction and calibration follow the standard processing routines for CHARA Classic data (as described in \citealt{ttb05} and \citealt{mca05}).  

For each calibrated observation, Table~1 lists the time of mid-exposure, the projected baseline $B$, the orientation of the baseline on the sky $\psi$, the visibility $V$, and the 1-$\sigma$ error to the visibility $\sigma V$ for each star.

The duplicity of these stars is not expected to affect our diameter measurements.  The secondary stars in these systems are all high contrast in the $K$-band, and our objects are considered as Hyades speckle singles in the infrared $K$ band according to \citet{pat98}.  These non-detections are not surprising.  For instance, $\delta^{1}$~Tau is a SB1 with a M-dwarf companion \citep{gri77} and  $\epsilon$~Tau is an exo-planet host star \citep{sat07}.  $\gamma$~Tau was resolved a single time as a speckle binary (with a large delta magnitude at $5000$\AA) by \citet{mor82}, having a system separation of 0.395 arcseconds.  Since this measurement, it has remained undetected as a binary by other programs \citep{mca78,mas93,pat98}.  In their infrared speckle program, \citet{pat98} did not detect a speckle companion for $\gamma$~Tau, but placed a limit to the $K$ band magnitude difference of $\Delta K=1.04$ for the system. We did not detect a separated fringe packet for the star in any of our observations, and hence we suggest that the detection from \citet{mor82} may be spurious.  The speckle binary, $\theta^{1}$~Tau, is also a SB1 (\citealt{tor97}, and references therein).  The companion to $\theta^{1}$~Tau, is a late F main sequence star \citep{pet81a}, which is supported by the non-detection in \citet{pat98}, where they list the limiting $\Delta K=4.6$ magnitudes.   As described in \citet{boy08}, our analysis of the binary $\mu$~Cas ($\Delta K=3.5$) shows that the interferometric diameter measured of the primary star of $\mu$~Cas is affected by $\sim1\%$ from the presence of the secondary.  Since the magnitude difference in $\theta^{1}$~Tau is at least one magnitude larger than this system, we neglect any possible influence the secondary might have on our visibility measurements of the primary star.

\placetable{tab1}      % Table 1 - Interferometric measurements of Hyades Giants

%%%%%%%%%%%%%%%%%%%%%%%%%%%%%%%%%%%%%%%%%%%%%%%%%%%%%%%%%%

\section{Angular Diameters and Stellar Parameters}        %Section 3

The uniform-disk $\theta_{\rm UD}$ and limb-darkened $\theta_{\rm LD}$ angular diameters are expressed as the following relations  
		%	EQUATION 1
\begin{equation}
V  =  \frac{2 J_1(\rm x) }{\rm x},
\end{equation}
		% EQUATION 2    
\begin{equation}
%\begin{split}
V = \left( {1-\mu_\lambda \over 2} + {\mu_\lambda \over 3} \right)^{-1} 
%\times \\
\times
\left[
(1-\mu_\lambda) {J_1(\rm x) \over \rm x} + \mu_\lambda {\left( \frac{\pi}{2} \right)^{1/2} \frac{J_{3/2}(\rm x)}{\rm x^{3/2}}} 
\right],
%\end{split}
\end{equation}
		%  EQUATION 3    
and \begin{equation}
\rm x = \pi \emph{B} \theta \lambda^{-1},
\end{equation}
where $\emph{J}_n$ is the $n^{th}$-order Bessel function, and $\mu_{\lambda}$ is the linear limb darkening coefficient at the wavelength of observation \footnote{In this work, we use $\mu_{K}=0.301$ for all Hyades giants \citep{cla95} }.  In Equation~3, $\emph{B}$ is the projected baseline in the sky, $\theta$ is the UD angular diameter of the star when applied to Equation~1 and the LD angular diameter when used in Equation~2, and $\lambda$ is the central wavelength of the observational bandpass \citep{han74}.  

We calculate the UD and LD diameter for each star from the calibrated visibilities by $\chi^2$ minimization of Equation~1 and Equation~2, where the error to the diameter fit is based upon the values on either side of the minimum for which $\chi^2$ = $\chi^2_{\rm min}$ + 1 \citep{pre92,wal03}.  Table~2 shows our results along with the reduced $\chi^2$ values for these diameter fits.  Note that our values for reduced $\chi^2$ are less than one, meaning we have overestimated the errors of the measured visibilities used in the diameter fits ($\sigma V$ in Table~1).  The best fits for the limb-darkened angular diameters to our calibrated visibilities and the 1-$\sigma$ errors are shown in Figure~1.  In our final analysis of $\epsilon$~Tau, we include the data point from \citet{van99}, which was taken at the same wavelength as our observations, in the fit.

The angular diameters of these stars are then transformed into linear radii $R$ using the \citet{van07} {\it Hipparcos} parallaxes.  In addition to these quantities, we calculate the effective temperature $T_{\rm eff}$ using the relation
        %    EQUATION 3
\begin{equation}
F_{\rm BOL} =  \frac{1}{4} \theta_{\rm LD}^2 \sigma T_{\rm eff}^4
\end{equation}
where $\sigma$ is the Stefan-Boltzmann constant.  

The bolometric flux $F_{\rm BOL}$ for each star was determined by applying the bolometric corrections of each star from \citet{all99}, assuming $M_{\rm BOL,\odot}$=4.74.  The results for the radius, bolometric flux, and effective temperature for each star are shown in Table~3. The significance of the luminosity sub-class IIIb for $\theta^{1}$~Tau (Table~2) is directly detected here in the smaller radius and $F_{\rm BOL}$ compared to the other giants.

\begin{figure}
\begin{center}
{\includegraphics[angle=90,height=12cm]{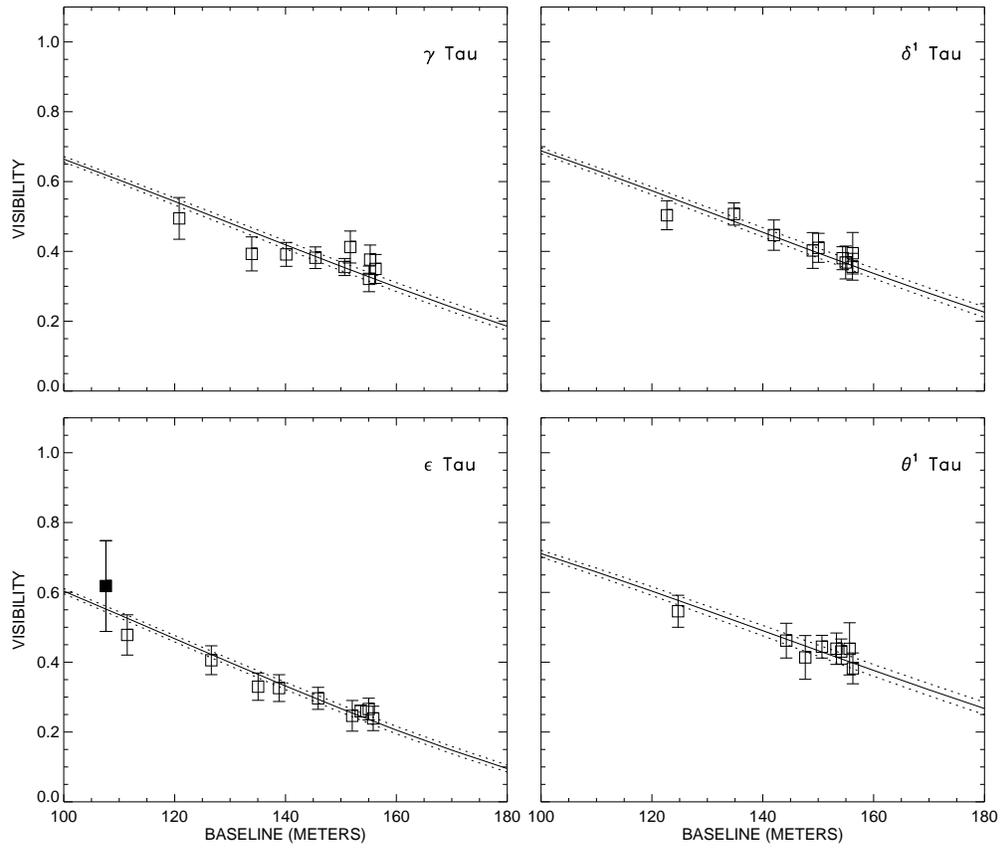}}
\end{center}
\caption{Limb darkened diameter fits to our data on the Hyades Giants.  The plot for $\epsilon$~Tau also shows the data point from \citet{van99} ({\it filled square}).}
\label{fig1}
\end{figure}

\placetable{tab2}        %Table 2-  Angular diameters of Hyades Giants
\placefigure{fig1}        % Figure1- LD diameter fit all giants  from Hyades/Targets/plothyades.pro
\placetable{tab3}        %Table 3-  Stellar properties of Hyades Giants

%%%%%%%%%%%%%%%%%%%%%%%%%%%%%%%%%%%%%%%%%%%%%%%%%%%%%%%%%%%%%%%%%%%%%%%

\section{Discussion}        % Section 4

Historically, each of these stars has been observed by lunar occultation (LO) and/or long baseline optical interferometry (LBOI) to obtain angular diameters (Table~4).  Diameters of three of the four giants have been measured by LO, and somewhat surprisingly, only two of the four giants had been measured by LBOI prior to this work.  While the LO measurements show a considerable scatter and large errors, the LBOI points also vary considerably within their errors with respect to each other.  Indeed, this is primarily an artifact of the relatively small size of these four stars creating quite a challenge for them to be sufficiently resolved with interferometers of modest baselines.  The advantages of observing stellar diameters with the long baselines of the CHARA Array are apparent, allowing us to obtain optimal sampling of the visibility curve.  For example, our measured diameter of $\epsilon$~Tau here includes the single PTI data point (Table~2, Figure~1), clearly improving the diameter fit from \citet{van99} (see Table~4).  Secondly, the sensitivity of our beam combiner allows us to observe calibrator stars that are very unresolved, closing the gap for systematic errors that may arise in the calibration process.  Additionally, these observations were made in the infrared, and are less subject to stellar limb darkening, making the transformations from the observed $\theta_{\rm UD}$ to the actual $\theta_{\rm LD}$ less model dependent.  

\placetable{tab4}        %Table 4-  Comparison of Angular Diameter Measurements of the Hyades Giants

For all existing angular diameter measurements from lunar occultation and long baseline optical interferometry (Table~4), we use Equation 1 to calculate the effective temperatures of these stars (Table~3, Direct Techniques).  For comparison, we show the range in effective temperature determinations when estimated via photometric and spectroscopic methods \citep{och00}, also in Table~3.  Our temperatures tend to lie on the low side of these ranges, which probably results from differences in model opacities and varying metallicity determinations of the models used in each reference.  In the case of $\theta^1$~Tau, the temperatures from spectroscopic techniques are higher than our derived temperature, which is likely to be an artifact of the duplicity of the star.

Figure~2 displays these available measurements on a H-R diagram for all stars, separated by the method of measurement.  To model these stars, we use the Padova database of stellar evolutionary tracks and isochrones\footnote{http://stev.oapd.inaf.it/cmd} \citep{mar07}, using a cluster turn-off age of 625~Myr \citep{per98}.  In Figure~2, we show isochrones for solar metallicity $Z_{\rm \odot}$=0.019 and two different metallicities of the Hyades cluster $Z_{\rm Hyades}$=0.024, 0.028 \citep{per98, the98}.  The model isochrone for both Hyades metallicities ($Z_{\rm Hyades}$=0.028, 0.024) are in excellent agreement with our observations.  To identify which part of this isochrone our stars were likely to lie, we investigated a single-star evolutionary track for a mass of 2.5$M_{\rm \odot}$ to determine which part of the isochrone a star would spend most of its lifetime \citep{gir00}.  We find that from the beginning of the core helium burning stage, up until the time helium is exhausted from the core, corresponds to $\sim$20\% of the star's total lifetime, second only to the time spent on the main-sequence, $\sim75$\% of its total lifetime.  The stars placement on Figure~2 clearly mark all four giants as residing on the helium burning Red Clump (RC), and this region is indicated as the thicker part of the Hyades metallicity isochrone of $Z_{\rm Hyades}$=0.028.  

Within this region of the RC, we look back to the model isochrones in order to determine a range of masses that these stars may have.  The model stellar mass for the lowest point of the RC is 2.48M$_{\rm \odot}$, and following this track up the end of the helium burning stage extends this model mass to 2.70M$_{\rm \odot}$.  These masses are consistent with the \citet{tor97} giant masses for the Hyades.  Assuming that these stars may fall anywhere between these masses, we predict a $\log g$ using the radii that we measure for each star (Table~3).  These values are in excellent  agreement with spectroscopically determined gravities found in the literature which have a large spread of values from $\log g$=2.2 to $\log g$=3.17, although for most estimates the gravity agrees with ours within 0.1~dex.   

\begin{figure}
\begin{center}
{\includegraphics[angle=0,height=12cm]{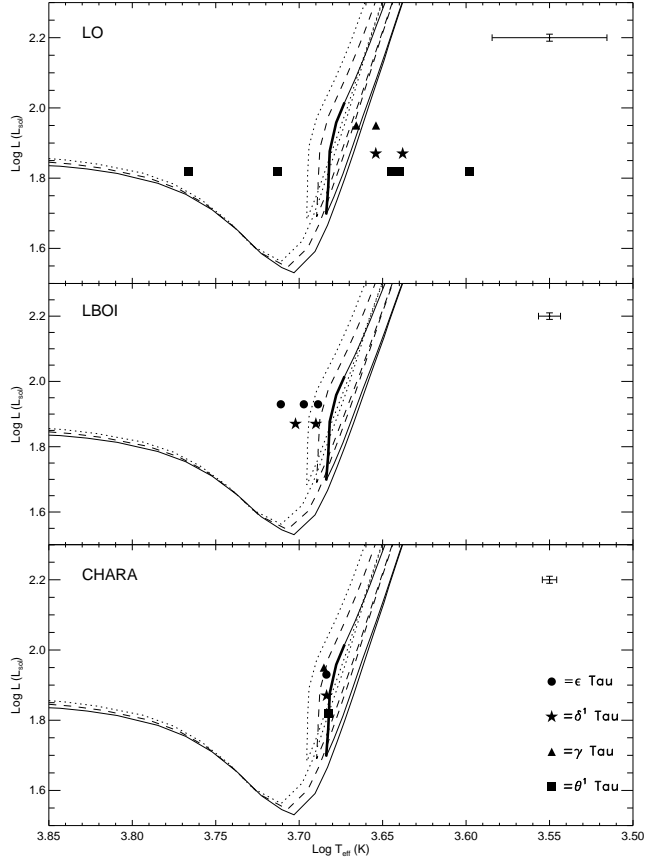}}
\end{center}
\caption{Effective temperatures derived from published angular diameter data from LO ({\it top panel}), previous LBOI ({\it middle panel}), and this work ({\it bottom panel}).  The symbols denoting the objects are consistent within all three panels, and the references for each measurements can be found in Table~4.  The typical 1-$\sigma$ error for each method is shown in the top right portion of each panel.  Padova model isochrones for 625~Myr are plotted for solar metallicity $Z_{\rm \odot}$=0.019 ({\it dotted line}) and metallicities $Z$=0.024 and $Z$=0.028 ({\it dashed line} and {\it solid line}, respectively).  The thick region of the Hyades isochrone for $Z=0.028$ identifies the region of the helium burning Red Clump.}
\label{fig2}
\end{figure}

\placefigure{fig2}        % Figure2 models and observations from /Hyades/Models/dartmod.pro

%%%%%%%%%%%%%%%%%%%%%%%%%%%%%%%%%%%%%%%%%%%%%%%%%%%%%%%%%%%%%%%

\acknowledgments

We would like to thank Gerard T. van Belle for his advice on the data anaysis.  The CHARA Array is funded by the National Science Foundation through NSF grant AST-0606958 and by Georgia State University through the College of Arts and Sciences. This research has made use of the SIMBAD literature database, operated at CDS, Strasbourg, France, and of NASA's Astrophysics Data System. This publication makes use of data products from the Two Micron All Sky Survey, which is a joint project of the University of Massachusetts and the Infrared Processing and Analysis Center/California Institute of Technology, funded by the National Aeronautics and Space Administration and the National Science Foundation.

%%%%%%%%%%%%%%%%%%%%%%%%%%%%%%%%%%%%%%%%%%%%%%%%%%%%%%%%%%%%%%%
% The bibliography starts here.
\clearpage
\bibliographystyle{apj}            % Please learn to use the
                                    % formatting of Latex's Bibtex. It
                                    % will make your life easier.
% apj.bst should be in this directory as well as apj-jour.bib and reference paper.bib
\bibliography{apj-jour,paper}      % "paper.bib" contains all my
                                    % references. "apj-jour.bib"
                                    % contains abbreviations of
                                    % journals.

%\clearpage
%
%%%%%%%%%%%%%%%%%%%%%%%%%%%%%%%%%%%%%%%%%%%%%%%%%%%%%%%%%%%%%%%%%%%%
% Tables

\clearpage
\begin{deluxetable}{lccccc}
\tablecolumns{10}
\tablenum{1}
\tabletypesize{\scriptsize}
\tablewidth{0pc} \tablecaption{Interferometric Measurements of Hyades Giants}
\tablehead{
\colhead{Star}    &
\colhead{JD} &  
\colhead{$B$} & 
\colhead{$\psi$}      &
\colhead{ } & 
\colhead{ } \\
\colhead{Name}    &
\colhead{($-$2,400,000)}  & 
\colhead{(m)} & 
\colhead{($^{\circ}$)}    & 
\colhead{$V$}          & 
\colhead{$\sigma V$}    }
\startdata
$\gamma$~Tau    &    54406.745    &    120.8    &    194.8    &    0.495    &    0.060    \\
$\gamma$~Tau    &    54406.770    &    133.9    &    195.6    &    0.393    &    0.049    \\
$\gamma$~Tau    &    54406.784    &    140.1    &    196.2    &    0.391    &    0.034    \\
$\gamma$~Tau    &    54406.799    &    145.4    &    197.0    &    0.382    &    0.031    \\
$\gamma$~Tau    &    54406.822    &    151.7    &    198.5    &    0.413    &    0.046    \\
$\gamma$~Tau    &    54406.842    &    155.1    &    200.0    &    0.321    &    0.037    \\
$\gamma$~Tau    &    54406.861    &    156.2    &    201.7    &    0.350    &    0.041    \\
$\gamma$~Tau    &    54406.884    &    155.3    &    204.1    &    0.377    &    0.042    \\
$\gamma$~Tau    &    54406.913    &    150.7    &    207.7    &    0.355    &    0.024    \\ \\
$\delta^1$~Tau    &    54406.752    &    122.7    &    193.4    &    0.504    &    0.041    \\
$\delta^1$~Tau    &    54406.776    &    134.8    &    194.7    &    0.507    &    0.032    \\
$\delta^1$~Tau    &    54406.793    &    142.0    &    195.7    &    0.447    &    0.043    \\
$\delta^1$~Tau    &    54406.819    &    150.1    &    197.6    &    0.410    &    0.042    \\
$\delta^1$~Tau    &    54406.846    &    155.0    &    199.9    &    0.368    &    0.047    \\
$\delta^1$~Tau    &    54406.865    &    156.2    &    201.8    &    0.395    &    0.059    \\
$\delta^1$~Tau    &    54406.874    &    156.2    &    202.8    &    0.356    &    0.038    \\
$\delta^1$~Tau    &    54406.897    &    154.4    &    205.6    &    0.380    &    0.033    \\
$\delta^1$~Tau    &    54406.925    &    149.0    &    209.8    &    0.403    &    0.051    \\ \\
$\epsilon$~Tau    &    54406.738    &    111.5    &    191.2    &    0.488    &    0.058    \\
$\epsilon$~Tau    &    54406.764    &    126.6    &    192.7    &    0.406    &    0.041    \\
$\epsilon$~Tau    &    54406.781    &    135.1    &    193.9    &    0.330    &    0.039    \\
$\epsilon$~Tau    &    54406.790    &    138.8    &    194.5    &    0.326    &    0.038    \\
$\epsilon$~Tau    &    54406.809    &    145.9    &    196.0    &    0.296    &    0.032    \\
$\epsilon$~Tau    &    54406.833    &    152.0    &    198.1    &    0.246    &    0.044    \\
$\epsilon$~Tau    &    54406.852    &    155.0    &    199.9    &    0.266    &    0.031    \\
$\epsilon$~Tau    &    54406.888    &    155.8    &    204.1    &    0.239    &    0.035    \\
$\epsilon$~Tau    &    54406.909    &    153.6    &    207.0    &    0.260    &    0.017    \\ \\
$\theta^1$~Tau    &    54406.758    &    124.7    &    194.7    &    0.546    &    0.046    \\
$\theta^1$~Tau    &    54406.802    &    144.2    &    196.7    &    0.462    &    0.050    \\
$\theta^1$~Tau    &    54406.812    &    147.7    &    197.3    &    0.414    &    0.063    \\
$\theta^1$~Tau    &    54406.836    &    153.3    &    199.0    &    0.439    &    0.045    \\
$\theta^1$~Tau    &    54406.855    &    155.7    &    200.6    &    0.438    &    0.075    \\
$\theta^1$~Tau    &    54406.871    &    156.3    &    202.1    &    0.382    &    0.044    \\
$\theta^1$~Tau    &    54406.900    &    154.2    &    205.3    &    0.430    &    0.036    \\
$\theta^1$~Tau    &    54406.919    &    150.7    &    207.9    &    0.444    &    0.033    \\
\enddata
\end{deluxetable}

%%%%%%%%%%%%%%%%%%%%%%%%%%%%%%%%%%%%%%%%%%%%%%%%%%%%%%%%%%%%%%%%%%%%                            

\newpage
% Table 2
\begin{deluxetable}{lcccccc}
\tabletypesize{\scriptsize}
\tablewidth{0pt}
\tablenum{2}
\tablecaption{Angular Diameters of Hyades Giants\label{tab2}}
\tablehead{
\colhead{Star}    &
\colhead{HR}    &
\colhead{Spectral Type}        &
\colhead{$\theta_{\rm UD}$}    &
\colhead{Reduced}    & 
\colhead{$\theta_{\rm LD}$}    &
\colhead{Reduced}    \\
\colhead{Name}    &
\colhead{}    &
\colhead{}    &
\colhead{(mas)}    &
\colhead{$\chi^{2}_{\rm UD}$}    &
\colhead{(mas)}    &
\colhead{$\chi^{2}_{\rm LD}$}    
}
\startdata
$\gamma$~Tau    &    HR~1346   &    K0~III    &    $2.452 \pm 0.033$    &    0.88    &    $2.517 \pm 0.034$    &    0.86    \\
$\delta^{1}$~Tau    &    HR~1373    &    K0~III    &    $2.347 \pm 0.037$    &    0.34    &    $2.408 \pm 0.038$    &    0.34    \\
$\epsilon$~Tau    &    HR~1409    &    G9.5~III    &    $2.660 \pm 0.030$    &    0.36    &    $2.734 \pm 0.031$    &    0.33    \\
$\epsilon$~Tau\tablenotemark{a}    &    HR~1409    &    G9.5~III    	&	$2.659 \pm 0.030$	&	0.33&	$2.733 \pm 0.031$&	0.32	\\
$\theta^{1}$~Tau    &    HR~1411    &    K0~IIIb    &    $2.247 \pm 0.042$    &    0.27    &    $2.305 \pm 0.043$    &    0.27    \\
\enddata
\tablenotetext{a}{Including \citet{van99} data point.}
\end{deluxetable}
\newpage
\clearpage
%
%%%%%%%%%%%%%%%%%%%%%%%%%%%%%%%%%%%%%%%%%%%%%%%%%%%%%%%%%%%%%%%%%%%%                            

% Table 3

\begin{deluxetable}{lcccccc}
\tabletypesize{\scriptsize}
\tablewidth{0pt}
\tablenum{3}
\tablecaption{Stellar Properties of Hyades Giants\label{tab3}}
\tablehead{
\colhead{Star}    &
\colhead{Radius}                &
\colhead{$\log g$\tablenotemark{a}}		&
\colhead{$F_{\rm BOL}$\tablenotemark{b}}					&
\colhead{$T_{\rm eff}$}                &
\colhead{Range of $T_{\rm eff}$ from}		&
\colhead{Range of $T_{\rm eff}$ from}			\\
\colhead{Name}    &
\colhead{($R_{\rm \odot}$)}    &
\colhead{(cgs)}				&
\colhead{(erg s$^{-1}$ cm$^{-2}$)}		&
\colhead{($K$)}    &
\colhead{Spectroscopy ($K$)}		&
\colhead{Direct Techniques\tablenotemark{c} ($K$)}		
}
\startdata
$\gamma$~Tau    &    $13.4 \pm 0.2$ \phn    &	2.58$-$2.61	&	$116\pm3$\phn\phn	&	$4844 \pm 47$\phn\phn    &    4800$-$4963    &	4508$-$4632	\\
$\delta^{1}$~Tau  &	$12.3 \pm 0.4$ \phn    &  2.65$-$2.69	&	$105\pm3$\phn\phn	&	$4826 \pm 51$\phn\phn    &    4750$-$5000    &	4335$-$5038	\\
$\epsilon$~Tau    &	$13.4 \pm 0.2$ \phn     &  2.59$-$2.63	&	$135\pm4$\phn\phn	&	$4827 \pm 44$\phn\phn    &    4656$-$4929     &	4883$-$5141	\\
$\theta^{1}$~Tau  &	$11.7 \pm 0.2$ \phn    &   2.69$-$2.73	&	$95\pm2$\phn		&	$4811 \pm 50$\phn\phn    &    4874$-$5000	  &	3962$-$5842	\\
\enddata
\tablenotetext{a}{Based upon mass range of 2.48$-$2.70 M$_{\rm \odot}$.}
\tablenotetext{b}{Expressed in $F_{\rm BOL}$/$1E-8$.  To correct for the light from the secondary component of $\theta^1$~Tau, a 3\% reduction to $F_{\rm BOL}$ was applied \citep{tor97, pet81b, pet81a}.} 
\tablenotetext{c}{Includes the LO and LBOI measured angular diameters, when available (see Table~4).}
\end{deluxetable}
\newpage
\clearpage

%%%%%%%%%%%%%%%%%%%%%%%%%%%%%%%%%%%%%%%%%%%%%%%%%%%%%%%%%%%%%%

\newpage
% Table 4
\begin{landscape}
\begin{deluxetable}{ccccccccc}
\tabletypesize{\scriptsize}
\tablewidth{0pt}
\tablenum{4}
\tablecaption{Comparison of Angular Diameter Measurements of the Hyades Giants\label{tab4}}
\tablehead{
\multicolumn{2}{c}{$\gamma$~Tau}    &
\multicolumn{2}{c}{$\delta^1$~Tau}    &
\multicolumn{2}{c}{$\epsilon$~Tau}    &
\multicolumn{2}{c}{$\theta^1$~Tau}    &
\colhead{Method,}    \\
\colhead{$\theta_{\rm LD} \pm \sigma$\phd\phd}    &
\colhead{$\Delta\theta_{\rm LD}$/$\sigma_{\rm C}$\tablenotemark{a}}    &
\colhead{$\theta_{\rm LD}\pm \sigma$\phd\phd}    &
\colhead{$\Delta\theta_{\rm LD}$/$\sigma_{\rm C}$\tablenotemark{a}}    &
\colhead{$\theta_{\rm LD}\pm \sigma$\phd\phd}    &
\colhead{$\Delta\theta_{\rm LD}$/$\sigma_{\rm C}$\tablenotemark{a}}    &
\colhead{$\theta_{\rm LD}\pm \sigma$\phd\phd}    &
\colhead{$\Delta\theta_{\rm LD}$/$\sigma_{\rm C}$\tablenotemark{a}}    &
\colhead{Reference\tablenotemark{b}}   
}
\startdata
$2.91 \pm 0.16$    &    $-2.4$\phs    &    \nodata	& \nodata	& \nodata	& \nodata	& \nodata	& \nodata	& LO, 1   \\
$2.75 \pm 0.18$	&	$-$1.3\phs	&	\nodata	& \nodata	& \nodata	& \nodata	& \nodata	& \nodata	& LO, 2	\\
\nodata	&	\nodata	& $2.97 \pm 0.7$\phn	&	$-0.8$\phs	& \nodata	& \nodata	& \nodata	& \nodata	& LO, 3   \\
\nodata	&	\nodata	& $2.76 \pm 0.7$\phn	&	$-0.5$\phs	& \nodata	& \nodata	& \nodata	& \nodata	& LO, 4   \\
\nodata	&	\nodata	&	\nodata	&	\nodata	& \nodata	& \nodata	&	$2.74 \pm 0.12$	& $-3.4$\phs	& LO, 5 \\
\nodata	&	\nodata	&	\nodata	&	\nodata	& \nodata	& \nodata	&	$1.56 \pm 0.45$	& $1.6$	& LO, 6 \\
\nodata	&	\nodata	&	\nodata	&	\nodata	& \nodata	& \nodata	&	$3.4 \pm 1.2$	& $-0.9$\phs	& LO, 7 \\
\nodata	&	\nodata	&	\nodata	&	\nodata	& \nodata	& \nodata	&	$2.0 \pm 0.2$	& $1.5$	& LO, 8 \\
\nodata	&	\nodata	&	\nodata	&	\nodata	& \nodata	& \nodata	&	$2.8 \pm 0.3$	& $-1.6$\phs	& LO, 9 \\
\nodata	&	\nodata	&	$2.338 \pm 0.033$	&	$1.4$ &	$2.671 \pm 0.032$	&	1.4 &	\nodata	&	\nodata	&	Mark III, 10	\\
\nodata	&	\nodata	&	$2.21 \pm 0.08$	&	2.2	&	$2.41 \pm 0.11$	&	2.8	&	\nodata	&	\nodata	&	NPOI, 11	\\
\nodata	&	\nodata	&	\nodata	&	\nodata	&	$2.57 \pm 0.06$	&	2.4	&	\nodata	&	\nodata	&	PTI, 12	\\
\hline
$2.517 \pm 0.034$	&	0.0	&	$2.408 \pm 0.038$	&	0.0	&	$2.733 \pm 0.031$	&	0.0	&	$2.305 \pm 0.043$	& 0.0	&	CHARA, this work	\\
\enddata
\tablenotetext{a}{Here, we define the combined error, $\sigma_{\rm C}=[\sigma_{\rm CHARA}^2+\sigma_{\rm Ref}^2]^{0.5}$, where $\sigma_{\rm Ref}$ is the error to the referenced measurement for each particular star entry.  $\Delta\theta_{\rm LD}$ is the difference between our angular diameter and the measurement for each reference.}
\tablenotetext{b}{1. \citet{rid80}, 2. \citet{ric98}, 3. \citet{kor84}, 4. \citet{tru87}, 5. \citet{rid82}, 6. \citet{rad80}, 7. \citet{bea82}, 8. \citet{eva81}, 9. \citet{whi79}, 10. \citet{moz03}, 11. \citet{nor01}, 12. \citet{van99}}
\end{deluxetable}
\end{landscape}
\newpage
\clearpage
%
%%%%%%%%%%%%%%%%%%%%%%%%%%%%%%%%%%%%%%%%%%%%%%%%%%%%%%%%%%%%%%%%%%%%                            

%                            Figure 1
\clearpage
%\input{epsf}
% Figure 1

%%%%%%%%%%%%%%%%%%%%%%%%%%%%%%%%%%%%%%%%%%%%%%%%%%%%%%%%%%%%%%
%                            Figure 2
\clearpage
%\input{epsf}
% Figure 2

%%%%%%%%%%%%%%%%%%%%%%%%%%%%%%%%%%%%%%%%%%%%%%%%%%%%%%%%%%%%%%

\end{document}